\documentclass[aps,prl,twocolumn,superscriptaddress,amsmath,amssymb,citeautoscript,floatfix]{revtex4-1}
\usepackage{graphicx}
\usepackage{bbm}
\usepackage{hyperref}
\usepackage[capitalize]{cleveref}
\usepackage{nicefrac}
\usepackage{float}
\usepackage{subfigure} 
\usepackage[shortcuts]{extdash}
\newcommand{\pf}{\operatorname{pf}}
\newcommand{\pdet}{\operatorname{pdet}}
\newcommand{\sech}{\operatorname{sech}}
\newcommand{\sgn}{\operatorname{sgn}}
\newcommand{\up}{{\uparrow}}
\newcommand{\down}{{\downarrow}}
\crefname{app}{Appendix}{Appendix}
\newcounter{app}\renewcommand{\theapp}{\Alph{app}}
\newcommand{\app}[1]{\refstepcounter{app}\subsection{Appendix \theapp:\, #1}\setcounter{equation}{0}\setcounter{figure}{0}}
\newcommand{\appendices}{\renewcommand{\section}{\app}\renewcommand{\theequation}{\Alph{app}\arabic{equation}}\renewcommand{\thefigure}{\Alph{app}\arabic{figure}}}
\begin{document}
\title{%
Signatures of topological phase transitions in Josephson current-phase discontinuities
}
\author{Pasquale Marra}
\email[]{pasquale.marra@spin.cnr.it}
\affiliation{CNR-SPIN, I-84084 Fisciano (Salerno), Italy}
\affiliation{Dipartimento di Fisica ``E. R. Caianiello'', Universit\`a degli Studi di Salerno, I-84084 Fisciano (Salerno), Italy}
\author{Roberta Citro}
\affiliation{CNR-SPIN, I-84084 Fisciano (Salerno), Italy}
\affiliation{Dipartimento di Fisica ``E. R. Caianiello'', Universit\`a degli Studi di Salerno, I-84084 Fisciano (Salerno), Italy}
\author{Alessandro Braggio}
\affiliation{CNR-SPIN, Via Dodecaneso 33, I-16146 Genova, Italy}
\affiliation{INFN Sezione di Genova, via Dodecaneso 33, I-16146, Genova, Italy}

\begin{abstract}

Topological superconductors differ from topologically trivial ones for the presence of topologically protected zero-energy modes.
To date, experimental evidence of topological superconductivity in nanostructures has been mainly obtained by measuring the zero-bias conductance peak via tunneling spectroscopy.
Here, we propose an alternative and complementary experimental recipe to detect topological phase transitions in these systems.
We show in fact that, for a finite-sized system with broken time-reversal symmetry, discontinuities in the Josephson current-phase relation correspond to the presence of zero-energy modes and to a change in the fermion parity of the groundstate.
Such discontinuities can be experimentally revealed by a characteristic temperature dependence of the current, and can be related to a finite anomalous current at zero phase in systems with broken phase-inversion symmetry.

\end{abstract}

\maketitle

\paragraph*{Introduction}
The recent discovery of topological materials deeply impacted condensed matter research\cite{Hasan2010_Zhang2011_Wehling2014}.
These materials exhibit a number of exceptional properties which are related to the presence of topologically protected states localized at their edges.
In topological superconductors (TS), for instance, Majorana edge states\cite{Kitaev2001} are characterized by a distinctive non-abelian statistics, which makes them ideal candidates for fault-tolerant quantum computation\cite{Kitaev20032_Nayak2008}.
Theoretically, a TS can be realized at the interface between a conventional superconductor and a topological insulator\cite{Fu2008}, or in semiconductor-superconductor heterostructures with spin-orbit coupling (SOC) in magnetic field\cite{Lutchyn2010_Oreg2010_Alicea2012_Rainis2014}.

However, revealing signatures of topological non-trivial phases in TS's is not straightforward\cite{Beenakker2013}.
This is mainly because, unlike more conventional continuous phase transitions, topological phase transitions\cite{Thouless1998} do not break any symmetry nor exhibit any critical behavior, but are instead identified by a change of the corresponding topological invariant\cite{Thouless1982_Tewari2012} and of the edge properties in continuous systems\cite{Zhou2008}.
To date, experimental evidence of non-trivial superconductivity comprises the measure of the $4\pi$-periodic Josephson current in TS rings\cite{Fu2009_Heck2011_Rokhinson2012_Wiedenmann2015}, or the zero-bias conductance peak via tunneling spectroscopy\cite{Mourik2012} or spatially-resolved spectroscopic imaging\cite{Nadj-Perge2014}.

\begin{figure}[t]
\centering
\includegraphics[scale=1,resolution=600]{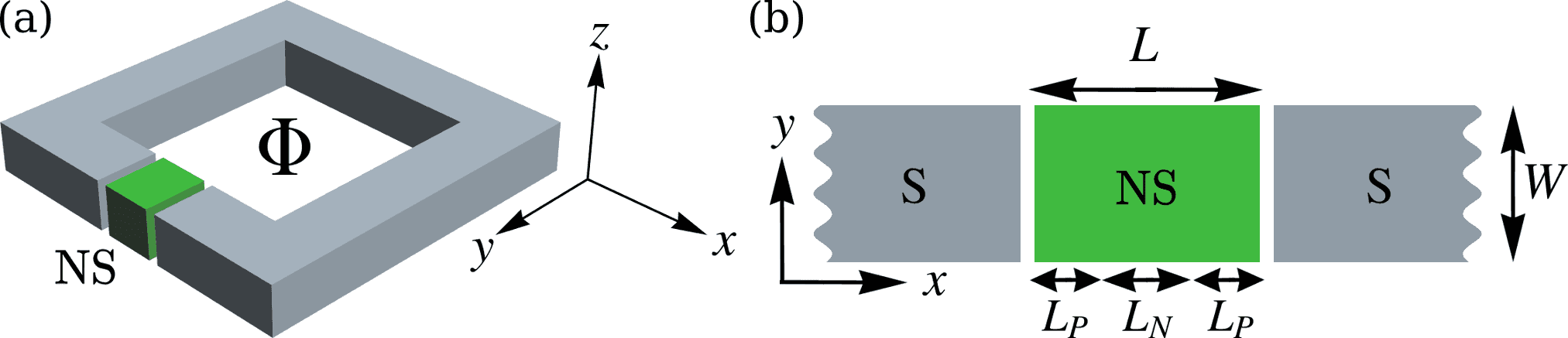}
\caption{
A TS (a) realized by a quantum nanostructure embedded into a superconducting ring with Rashba SOC $\propto(\boldsymbol{\sigma}\times\mathbf{p})_z$ along the $z$ axis.
The nanostructure (b) of length $L$, width $W$, and proximity length $L_P$ ($L_N=L-2L_P$), connected to two superconducting leads, can be either a quantum dot ($L=W=1$ lattice site), wire ($L>W=1$), or planar well ($L, W>1$).
}%
\label{fig:System}
\end{figure}

In this work 
we propose a direct method to probe variations of the topological invariant, i.e., the fermion parity of the groundstate\cite{Zocher2013_Budich2013}, 
in Josephson junctions\cite{Beenakker2013PRL,Diez2013_Barbarino2013_San-Jose2014}.
It is known that the Josephson current-phase relation (CPR) may exhibit discontinuities at low temperatures corresponding to Andreev level crossings\cite{Balatsky2006}, in systems as different as quantum dots\cite{Cheng2009_Brunetti2013_Kirsanskas2015}, nanowires\cite{Yokohama2013}, Josephson junction arrays\cite{Koch2008}, and Weyl semimetals\cite{Khanna2015}.
Here, we show that in finite-sized non-interacting $s$-wave TS's with broken 
time-reversal symmetry, 
such discontinuities are related to the presence of zero-energy modes and to variations of the fermion parity, which define the topological phase transitions of this system.
These discontinuities can be experimentally revealed by a characteristic temperature dependence and are moreover related, in systems with broken phase-inversion symmetry, to an anomalous current at zero phase.

\paragraph*{The model}
A TS can be realized by a nanostructure embedded into a superconducting ring\cite{Giazotto2011} with SOC and magnetic field\cite{Lutchyn2010_Oreg2010_Alicea2012_Rainis2014} (see \cref{fig:System}).
Here, we consider the case of finite-sized systems, e.g., a quantum dot, wire, or planar well 
(see \cref{app:Hamiltonian}), 
described by a tight-binding Bogoliubov-de~Gennes (BdG) Hamiltonian\cite{deGennes1999} 
with linear dimensions smaller than the coherence length $\xi=\hbar v_F/\Delta$, 
where $v_F$ is the Fermi velocity and $\Delta$ the superconducting gap.
This system is the zero-dimensional (0D) limit of a one-dimensional (1D) Majorana chain\cite{Kitaev2001}, in the sense that its Hamiltonian does not depend on any momentum-like continuous parameter.
The Andreev spectrum of this system is a discrete set of particle-hole energy levels $E_i(\varphi)$ which depend on the gauge-invariant phase difference $\varphi$ of the superconducting order parameter between the two leads, induced by a magnetic flux $\Phi$.
If $L\ll\xi$, where $L$ is the distance between the leads, the CPR is given by\cite{Golubov2004_Nazarov2009Quantum}
\begin{equation}
I(\varphi)=\frac{e}{\hbar}\sum_{i} f\left[\frac{E_i(\varphi)}{k_B T}\right] \partial_\varphi E_i(\varphi),
\label{eq:Josephson}
\end{equation}
where $f(x)=1/(e^x+1)$ is the Fermi-Dirac distribution.
Note that the low-energy Andreev spectrum does not depend on the superconducting ring length, in the limit of short junctions\cite{Affleck2000} 
(see \cref{app:Hamiltonian}).
If the lowest energy (LE) levels close the gap with linear phase-dispersion, the CPR exhibits a discontinuity at zero temperature $T=0$.
In fact, in this case the Fermi-Dirac distribution in \cref{eq:Josephson} converges to a step function, and thus the only contribution to the current is given by energy levels $E\le 0$, i.e., $I(\varphi)=(e/\hbar)\sum_{E_i\le0}\partial_\varphi E_i(\varphi)$.
This mandates a discontinuous drop $\Delta I(\varphi^*)=-(e/\hbar)\sum_{E_j(\varphi^*)=0}|\partial_\varphi E_j(\varphi^*)|$ at 
any gapless point $\varphi^*$ where zero-energy levels $E_j(\varphi^*)=0$ have linear phase-dispersion $\partial_\varphi E_j(\varphi^*)\neq0$.
Since Andreev levels are continuously differentiable in finite-sized systems 
(see \cref{app:Hamiltonian} 
and Refs.~\onlinecite{Bronshtein1979_Colombini2012}) these are the only points where the CPR can be discontinuous.
Hence, discontinuities at zero temperature correspond to zero-energy modes closing the particle-hole gap.
In general, the converse is not true, i.e., the gap may close without any discontinuity if $\partial_\varphi E_j(\varphi^*)=0$.
Hereafter we will show that such discontinuities correspond,
if time-reversal symmetry is broken,
to fermion parity transitions
and thus can be related to a topological phase transition of the TS\@.

\begin{figure}
\centering
\includegraphics[scale=1,resolution=600]{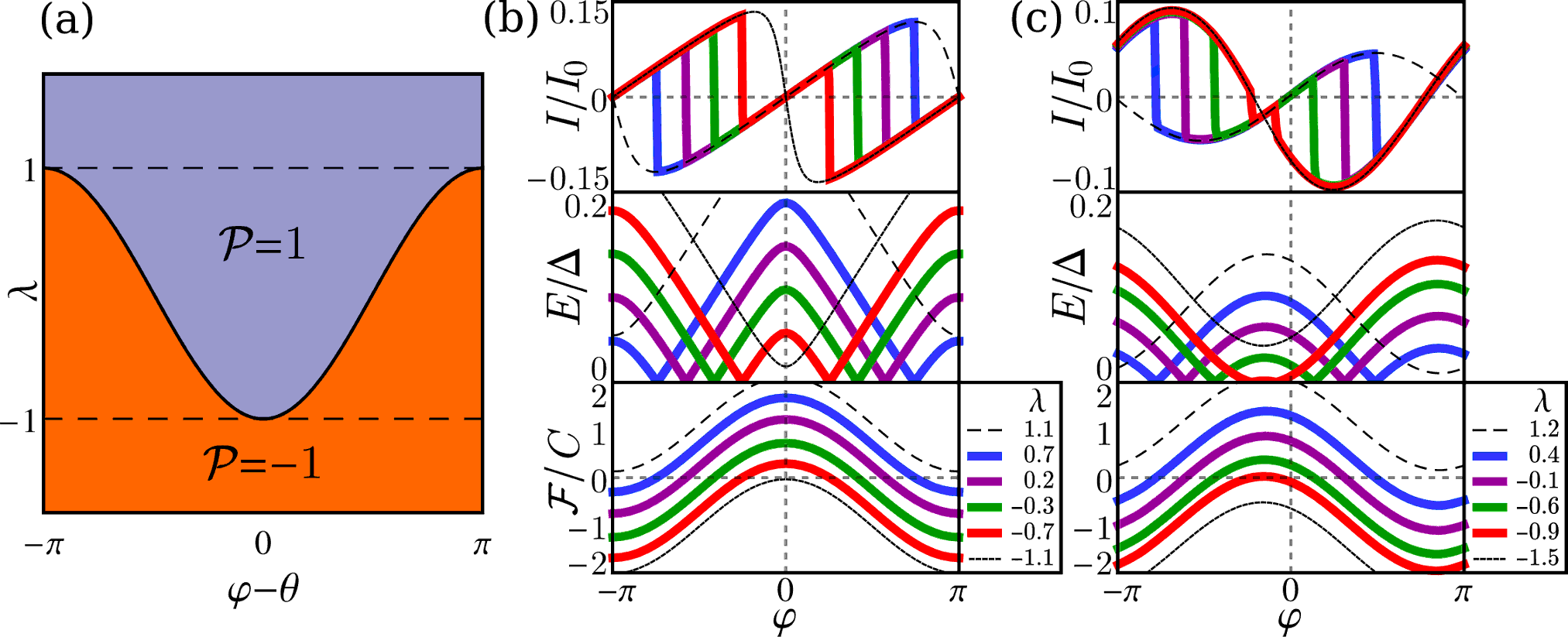}
\subfigure{\label{fig:PhaseSpace}}
\subfigure{\label{fig:1D}}
\subfigure{\label{fig:2D}}
\vspace{-12pt}
\caption{
Topological phase space (a) of a finite-sized TS as a function of $\varphi$, $\theta$ and $\lambda$.
Topological transitions between states with different fermion parity are possible for $|\lambda|<1$, where 
$\mathcal{P}_\varphi=\pm1$
respectively for $|\varphi-\theta|\lessgtr\arccos\left(-\lambda\right)$.
Zero-energy modes occur at $\cos(\varphi-\theta)=-\lambda$ (solid line).
CPR in units of $I_0=(e/\hbar)\Delta$,
LE Andreev level, and 
Pfaffian $\mathcal{F}_\varphi/C$ for a quantum wire (b) and a planar well (c) with magnetic field $b=b_y$ perpendicular to the SOC and current directions.
Different magnetic field directions give qualitatively similar results.
For $|\lambda|<1$ (continuous lines), CPR discontinuities correspond to zero-energy modes ($\mathcal{F}_{\varphi^*}=0$) and to variations of the fermion parity 
$\mathcal{P}_\varphi=\sgn \mathcal{F}_\varphi$ 
at $\varphi^*_\pm=\theta\pm\arccos(-\lambda)$.
The trivial $\mathcal{F}_\varphi>0$ and non-trivial $\mathcal{F}_\varphi<0$ branches of the CPR correspond respectively to $\lambda>1$ (dashed lines) and $\lambda<-1$ (dotted lines).
In a planar well with $b_y\neq0$ and $\alpha\neq0$ (c), the CPR and Pfaffian are no longer symmetric under phase-inversion $\varphi\rightarrow-\varphi$ ($\theta\neq n\pi$ in this case), 
and an anomalous current $I(0)\neq0$ is present for $\lambda=-0.9$ and $-1.5$.
}%
\label{fig:PhaseSpacePfaffianJosephson}
\end{figure}

Superconductors exhibit particle-hole symmetry, i.e., their BdG Hamiltonian $\mathcal{H}(\varphi)$ is invariant under the antiunitary transformation $\Xi=\tau_x K$ with $\Xi^2=1$, where $K$ is the complex conjugate operator and $\tau_x$ the Pauli matrix in particle-hole space.
For a finite magnetic field $\mathbf{b}\neq0$, 
time-reversal and chiral symmetries 
are broken, and hence the system is in the Altland-Zirnbauer\cite{Altland1997_Schnyder2009_Kitaev2009} symmetry class D.
This class is characterized both in 1D (Majorana chain, continuous spectrum) and in 0D (finite-sized system with discrete energy spectrum) by the $\mathbb{Z}_2$ topological invariant 
accordingly to the periodic table of topological phases\cite{Altland1997_Schnyder2009_Kitaev2009}.
Analogously to the 1D case (Majorana chain), the topological invariant in a 0D system is defined
following Ref.~\onlinecite{Loring2015} 
as the fermion parity of the groundstate\cite{Kitaev2001,Zocher2013_Budich2013} 
$\mathcal{P}_\varphi=\sgn\mathcal{F}_\varphi$, 
where $\mathcal{F}_\varphi=\pf\left[\mathcal{H}(\varphi)\imath\tau_x\right]$ is the Pfaffian\cite{Wimmer2012} of the matrix $\mathcal{H}(\varphi)\imath\tau_x$.
The fermion parity labels the topological inequivalent phases, i.e., trivial 
$\mathcal{P}_\varphi=1$
($\mathcal{F}_\varphi>0$) and non-trivial 
$\mathcal{P}_\varphi=-1$
($\mathcal{F}_\varphi<0$).
Moreover, since $\mathcal{F}_\varphi^2=\det[\mathcal{H}(\varphi)\imath\tau_x]=\det[\mathcal{H}(\varphi)]=\prod_i E_i(\varphi)$, the condition $\mathcal{F}_{\varphi^*}=0$ corresponds to gapless points $\varphi^*$ where zero-energy modes occur.
The topological invariant in 0D (finite-sized system) and in 1D (continuous limit) are closely related:
in the limit $L\rightarrow\infty$ in fact, 
the Majorana number\cite{Kitaev2001} coincides with the fermion parity of the groundstate, i.e., $\mathcal{M}=\sgn\{\pf[{\mathcal{H}\imath\tau_x}]\}$\cite{Zocher2013_Budich2013,Loring2015}.

\paragraph*{Fermion parity phase-dependence}
The non-trivial phase of a Majorana chain (1D, continuous spectrum) requires an open particle-hole gap, which can be realized only with SOC\cite{Lutchyn2010_Oreg2010_Alicea2012_Rainis2014} and for specific magnetic field directions\cite{Lin2012_Osca2014_Rex2014}.
Contrarily, in 0D finite-sized systems, the topological invariant 
$\mathcal{P}_\varphi$
is well defined even in absence of SOC and for any magnetic field direction, as long as $\mathcal{F}_\varphi\neq0$, and depends explicitly on the phase $\varphi$.
Changes in the fermion parity 
$\mathcal{P}_\varphi$
define the topological phase transitions in this system.
Independently from the details of the tight-binding Hamiltonian, the Pfaffian can be expanded as a Fourier series in the phase $\varphi$ with coefficients $a_n\propto\Delta^{2n}$.
If the superconducting gap is smaller than the bandwidth of the nanostructure (e.g., in conventional superconductors), higher harmonics become negligible and one obtains 
at the first order 
(see \cref{app:Pfaffian}) 
\begin{align}
\mathcal{F}_\varphi\approx C\left[ \cos(\varphi-\theta)+\lambda\right],
\label{eq:PfaffianLimit}
\end{align}
where 
$\lambda=(\mathcal{F}_{\varphi'}+\mathcal{F}_{\varphi'+\pi})/(2C)$ (the sum does not depend on the choice of the angle $\varphi'$), $2C=\sqrt{(\mathcal{F}_{0}-\mathcal{F}_{\pi})^2+(\mathcal{F}_{\nicefrac{\pi}{2}}-\mathcal{F}_{-\nicefrac{\pi}{2}})^2}$, and $\tan\theta=(\mathcal{F}_{\nicefrac{\pi}{2}}-\mathcal{F}_{-\nicefrac{\pi}{2}})/(\mathcal{F}_{0}-\mathcal{F}_{\pi})$.
These parameters depend on the Hamiltonian, e.g., on the hopping parameter $t$, magnetic field $\mathbf{b}$, chemical potential $\mu$, SOC $\alpha$, and superconducting gap $\Delta$, but not on the gauge-invariant phase $\varphi$.
The topological phase space is thus completely characterized by the phase $\varphi$ and the parameter $\lambda=\lambda(t,\alpha,\mu,\mathbf{b},\Delta)$, as shown in 
\cref{fig:PhaseSpace}.
If $|\lambda|\ge1$ the system is either in the trivial $\mathcal{F}_\varphi>0$ for $\lambda\ge1$ or non-trivial phase $\mathcal{F}_\varphi<0$ for $\lambda\le-1$, with the exception of a single gapless point at $\varphi^*=\theta+\pi$ or $\theta$ respectively for $\lambda=\pm1$.
If $|\lambda|<1$ instead, topological transitions occur at the gapless points $\varphi^*_\pm\approx\theta\pm\arccos(-\lambda)$.
Hence, the closing of the particle-hole gap $\mathcal{F}_\varphi=0$ defines the boundaries between trivial and non-trivial phases.

Furthermore, if SOC or magnetic field vanish, the Andreev spectrum, CPR, and fermion parity are invariant under phase-inversion, 
which mandates $\theta=n\pi$ with $n$ integer in \cref{eq:PfaffianLimit}.
This invariance is related to the magnetic mirror symmetry\cite{Lu2015} $M_{xz}\Theta=K$, where $M_{xz}=\imath\sigma_y$ is the spin mirror reflection across the $xz$ plane and $\Theta=-\imath\sigma_y K$ the time-reversal operator.
If $\alpha=0$ or $\mathbf{b}=0$ in fact, the quantization axis can be arbitrarily chosen such that the only complex terms in the BdG Hamiltonian are those in the phase $\varphi$ 
(see \cref{app:Pfaffian}).
Therefore $\mathcal{H}(\varphi)^*=\mathcal{H}(-\varphi)$, and consequently $I(\varphi)=-I(-\varphi)$, 
$\mathcal{P}_\varphi=\mathcal{P}_{-\varphi}$, 
and $\theta= n\pi$.
However, if the magnetic mirror symmetry is broken, i.e., if $\mathcal{H}(\varphi)^*\neq\mathcal{H}(-\varphi)$, the CPR and the fermion parity may be no longer symmetric under phase-inversion, i.e., $I(\varphi)\neq-I(-\varphi)$, 
$\mathcal{P}_\varphi\neq\mathcal{P}_{-\varphi}$, 
and $\theta\neq n\pi$.
In this case, the magnetic mirror symmetry corresponds to the inversion of the Pfaffian phase-shift $\theta\rightarrow-\theta$ 
(see \cref{app:Pfaffian}), 
and the fermion parity is still invariant under the more general transformation $\varphi\rightarrow2\theta-\varphi$, i.e., 
$\mathcal{P}_\varphi=\mathcal{P}_{2\theta-\varphi}$
[cf.~\cref{eq:PfaffianLimit}].
A Pfaffian phase-shift $\theta\neq n\pi$ is thus a signature of the broken phase-inversion and magnetic mirror symmetries, and can result in an anomalous current\cite{Kulagina2014_Yokoyama2014_Campagnano2015,Dolcini2015,Bergeret2015} at zero phase
(see below).

\paragraph*{CPR discontinuities}
In a superconductor with broken 
time-reversal symmetry, 
any zero-energy mode is at least doubly-degenerate due to particle-hole symmetry.
Hence the Pfaffian can be expanded near any gapless point as $\mathcal{F}_\varphi\propto(\varphi-\varphi^*)^d$ where the order $d$ is half the total multiplicity $2d=\sum_j m_j$, with $m_j$ the multiplicities of zero-energy modes $E_j(\varphi)\propto(\varphi-\varphi^*)^{m_j}$.
Thus, if the Pfaffian first derivative $\mathcal{F}_{\varphi}^{\prime}=\partial_\varphi\mathcal{F}_{\varphi}$ is non-zero at the gapless point $\varphi^*$, there exist only two doubly-degenerate LE levels $E_\pm(\varphi^*)=0$ with linear phase-dispersion $E_\pm(\varphi)\propto(\varphi-\varphi^*)$.
Therefore one obtains $\partial_\varphi E_\pm(\varphi^*)=\pm\mathcal{F}_{\varphi^*}^{\prime}/\chi_{\varphi^*}$, where 
$\chi_{\varphi^*}=\prod_{E_i>0}E_i(\varphi^*)>0$ 
(see \cref{app:Energy}).
Hence, the LE level contribution changes its sign passing through the gapless point, and the total current exhibits a discontinuous drop given by
\begin{equation}
\Delta I\left(\varphi^*\right)
=-
\frac{2e}\hbar
\frac{|\mathcal{F}_{\varphi^*}^{\prime}|}
{{\chi_{\varphi^*}}}
,
\label{eq:Discontinuity}
\end{equation}
where $|\mathcal{F}_{\varphi^*}^{\prime}|=C|\sin{(\varphi^*-\theta)}|=C\sqrt{1-\lambda^2}$ [cf.~\cref{eq:PfaffianLimit}].
\Cref{eq:Discontinuity} relates CPR discontinuities at zero temperature with the variations of the fermion parity 
$\mathcal{P}_\varphi=\sgn\mathcal{F}_\varphi$
in superconductors with broken 
time-reversal symmetry, 
and is valid for $\mathcal{F}_{\varphi^*}^{\prime}\neq0$.
If $|\lambda|<1$ in fact, the Pfaffian changes its sign at $\varphi^*_\pm=\theta\pm\arccos(-\lambda)$, where $\mathcal{F}'_{\varphi^*_\pm}\neq0$ according to \cref{eq:PfaffianLimit}.
Here, the CPR has two discontinuities $\Delta I(\varphi_\pm^*)\neq0$ which correspond to transitions between topological phases with even and odd fermion parity, where zero-energy modes appear.
If $|\lambda|>1$ instead, no zero-energy mode nor CPR discontinuity occur.
In the limit cases $\lambda=\pm1$, the Pfaffian vanishes at the gapless point without any sign-change ($\mathcal{F}_{\varphi^*}^{\prime}=0$), while the CPR may exhibit at most one discontinuity, corresponding to a zero-energy mode with $m_\pm=1$.
Hence, 
the presence of two distinct discontinuities in the CPR of a TS with broken 
time-reversal symmetry 
defines the boundaries between inequivalent topological phases.
Moreover, these discontinuities are a direct signature of zero-energy modes, and coincide with a sign-change of the LE level contribution to the current.
These zero-energy modes signal the topological transition at $\varphi^*$ between phases with different fermion parity, and are described locally as a linear superposition of particle and hole states or, equivalently, of two orthogonal Majorana states 
(see \cref{app:Majorana}).
The results presented here hold for \emph{any} discrete 0D Hamiltonian in the Altland-Zirnbauer class D (particle-hole symmetry and broken 
time-reversal symmetry).
Note that, if 
time-reversal symmetry 
is unbroken ($\mathbf{b}=0$ and $\theta=n\pi$), the total multiplicity is $2d\ge4$ due to spin and particle-hole degeneracy, and therefore $\mathcal{F}_{\varphi^*}^{\prime}=0$ at any gapless point.
Hence, no fermion parity transition occurs ($|\lambda|>1$) and the CPR may exhibit at most one discontinuity at $\varphi^*=n\pi$ if $|\lambda|=1$, e.g., in point contact junctions\cite{Golubov2004_Nazarov2009Quantum}.
Indeed, a time-reversal invariant $s$-wave superconductor is topologically trivial.
Since CPR discontinuities correspond to an abrupt jump to the LE state, they can be measured at the equilibrium (DC Josephson current) and are not affected by quasiparticle poisoning\cite{Rainis2012}.

\paragraph*{Numerical results}

\Cref{fig:1D,fig:2D} 
show the CPR at zero temperature, the LE Andreev level, and the Pfaffian of a quantum wire ($W=1$) and a planar well ($W>1$) as a function of the phase and magnetic field $b=b_y$ perpendicular to the SOC [$\propto(\boldsymbol{\sigma}\times\mathbf{p})_z$] and current directions, calculated directly from the BdG Hamiltonian $\mathcal{H}(\varphi)$ 
(see \cref{app:Hamiltonian} for details).
For $|\lambda|<1$, the gap closes with linear phase-dispersion 
and thus the CPR has two discontinuities 
at 
$\varphi^*_\pm$.
In a planar well with $b_y\neq0$ and $\alpha\neq0$ 
[\cref{fig:2D}], 
the CPR and the Pfaffian are no longer symmetric under phase-inversion $\varphi\rightarrow-\varphi$ ($\theta\neq n\pi$ in this case).
Note that fermion parity transitions and CPR discontinuities are present for any magnetic field direction.
In a quantum wire with $L\rightarrow\infty$ (1D, continuous spectrum), 
we verified numerically that \cref{eq:PfaffianLimit,eq:Discontinuity} reproduce the well-known results of Refs.~\onlinecite{Lutchyn2010_Oreg2010_Alicea2012_Rainis2014}.
In particular, the trivial 
($\mathcal{M}=1$)
and non-trivial 
($\mathcal{M}=-1$)
phases
correspond respectively to 
$\lambda>1$ and $\lambda=-1$ in \cref{eq:PfaffianLimit}.
We have also verified 
that Majorana bound states localize at the wire edges, corresponding to
a discontinuity in the CPR at $\varphi^*=\pi$ if $\mathbf{b}\perp y$.

\paragraph*{Anomalous current}

When the magnetic mirror symmetry is broken ($\theta\neq n\pi$), the current may be no longer symmetric under phase-inversion, which can result in a finite anomalous current\cite{Kulagina2014_Yokoyama2014_Campagnano2015,Dolcini2015,Bergeret2015} at $\varphi=0$, as shown in 
\cref{fig:2D}.
This can be realized, e.g., in planar quantum wells with SOC and magnetic field, where a finite anomalous current has been related to the presence of chiral edge states\cite{Dolcini2015} or to a SOC-induced Lorentz force\cite{Bergeret2015}.
In these systems, 
a topological phase transition can be
revealed by a 
discontinuity of the current at zero phase $\Delta I(\varphi=0,\nu)$ with respect to the parameter $\nu$ (magnetic field, chemical potential, or SOC) which drives the system through the topological transition.
In this case one can always find a value $\nu=\nu^*$ such that the gap closes at $\varphi^*=0$, i.e., $\lambda(\nu^*)=-\cos\theta$ 
[cf. \cref{eq:PfaffianLimit,fig:PhaseSpace}].
At this gapless point the fermion parity changes 
and the current at $\varphi=0$ has a 
discontinuous drop with respect to the parameter $\nu$ given by
\begin{equation}
\Delta I(\varphi^*=0,\nu=\nu^*)=-\frac{2e}\hbar \frac{C}{\chi_0}\left\vert\sin{\theta}\right\vert.
\label{eq:Anomalous}
\end{equation}
The value $\nu^*$ corresponds 
to a crossover between the trivial $\mathcal{F}_\varphi>0$ ($\lambda>1$) and non-trivial $\mathcal{F}_\varphi<0$ ($\lambda<-1$) branches of the CPR, as shown in 
\cref{fig:2D}.
This discontinuity mandates a finite anomalous current $I(0,\nu)\neq0$ near $\nu=\nu^*$ either in the trivial or non-trivial phases, or in both.
Numerical calculations indicate that it is non-zero only in the non-trivial phase 
[cf.~\cref{fig:2D}].
Hence, a discontinuity of the anomalous current at zero phase with respect to any system parameter (e.g., magnetic field) is also a signature of a topological phase transition.

\begin{figure}
\centering
\includegraphics[scale=1,resolution=600]{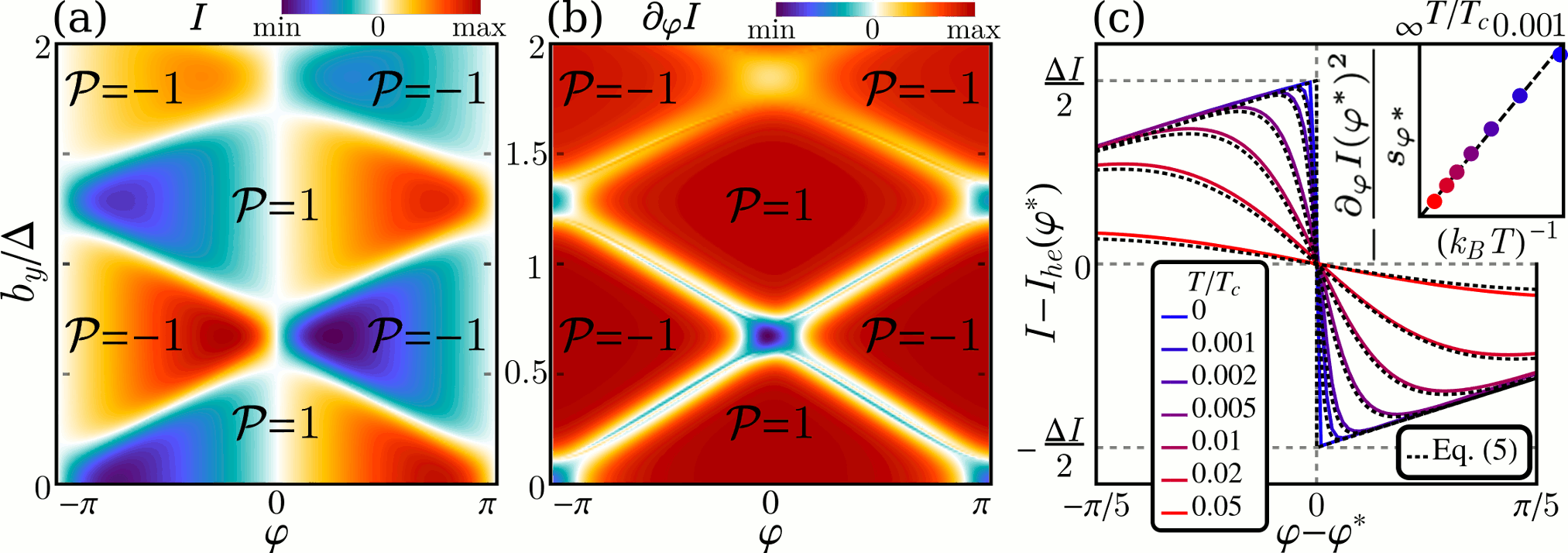}
\subfigure{\label{fig:JosephsonBT}}
\subfigure{\label{fig:Derivative}}
\subfigure{\label{fig:JosephsonSmooth}}
\vspace{-12pt}
\caption{
Fermion parity 
$\mathcal{P}_\varphi$, 
CPR (a), and phase-derivative $\partial_\varphi I(\varphi)$ of the current (b) at $T=0.02\,T_c$ of a quantum wire as a function of the magnetic field along the $y$ axis.
Spikes in the phase-derivative correspond to topological phase transitions.
Josephson current (c) near the gapless point and phase-derivative $\partial_\varphi I(\varphi^*)$ (inset).
Dotted lines correspond to \cref{eq:JosephsonSmooth}, continuous lines to numerical calculations.
The two minima (b) at $\varphi=0$ 
($b_y/\Delta\approx0.5$ and $\approx2$ respectively)
are constant in temperature, i.e., do not diverge for $T\rightarrow 0$, and indeed do not correspond to any topological phase transition.
}%
\label{fig:JosephsonDerivative}
\end{figure}

\paragraph*{Experimental proposal}
At finite temperatures, CPR discontinuities are smoothed out by the thermal spreading of the Fermi-Dirac distribution.
For $T\ll T_d=\delta_d/k_B$, where $\delta_d$ is the gap between the first and second Andreev levels at $\varphi^*$, 
the current can be expanded as a sum of two contributions $I_{he}(\varphi)$ and $I_{le}(\varphi)$ coming respectively from higher energy levels ($E>\delta_d$) and from the LE level.
The latter contribution depends strongly on the temperature and can be obtained from \cref{eq:Josephson,eq:Discontinuity} 
(see \cref{app:Temperature}), 
which yield
\begin{equation}
I(\varphi)
\!\approx\! I_{he}(\varphi)
+\frac{\Delta I(\varphi^*)}2
\tanh\!{\left[-\frac{\hbar}{e}\frac{\Delta I(\varphi^*)(\varphi-\varphi^*)}{4 k_B T}\right]}.
\label{eq:JosephsonSmooth}
\end{equation}
The current phase-derivative diverges as $\partial_\varphi I(\varphi^*)\approx-\hbar/(8e)\Delta I(\varphi^*)^2/(k_B T)$ for $T\rightarrow0$.
Hence, the scaling factor $s_{\varphi^*}=-\partial_\varphi I(\varphi^*) k_B T$ is a direct measure of the discontinuous drop, since $\Delta I(\varphi^*)\approx -\sqrt{(8e/\hbar) s_{\varphi^*}}$.
Since the separation between energy levels increases as the system size decreases, the temperature $T_d$, below which CPR discontinuities are measurable, 
is maximized for small linear dimensions.
For parameters considered in 
\cref{fig:PhaseSpacePfaffianJosephson,fig:JosephsonDerivative}, 
we obtain $T_d\approx0.2\,T_c$.

In light of this, we propose to measure the low-temperature CPR\cite{Golubov2004_Nazarov2009Quantum,Frolov2004_Sochnikov2013_Szombati2015} through a nanostructure with 
broken 
time-reversal symmetry.
\Cref{fig:JosephsonBT,fig:Derivative} 
show the CPR of a quantum wire ($L=200$ sites) at $T>0$ and its phase-derivative as a function of the magnetic field.
Besides, 
\cref{fig:JosephsonSmooth} 
shows the effect of temperature on current discontinuities, which is described by \cref{eq:JosephsonSmooth}.
Spikes in the phase-derivative which exhibit the characteristic temperature scaling of 
\cref{fig:JosephsonSmooth} 
identify the boundaries between inequivalent topological phases.
These effects should be measurable in clean Josephson weak-links of the order of 50--150~nm in InAs or InSb nanowires ($L=100$--300 lattice sites) proximized by a conventional superconductor\cite{Giazotto2011} (e.g., Nb or Al).
Note that the Josephson critical current has been recently measured in an InAs nanowire in the relevant regime\cite{Giazotto2016}, at very low temperatures $\sim 10~\mathrm{mK}$ comparable with the temperatures considered here. 

\paragraph*{Conclusions}
We have shown that discontinuities in the Josephson CPR correspond, in TS's with broken time-reversal symmetry, to topological phase transitions between states with different fermion parity, where zero-energy modes occur.
These current discontinuities are not affected by quasiparticle poisoning and can be revealed by spikes in the current phase-derivative and by their characteristic temperature dependence.
Moreover, in systems with broken phase-inversion symmetry, topological phase transitions correspond to discontinuities of the anomalous current at zero phase.
Such features in the CPR provide an experimental tool to probe the fermion parity and to resolve the topological phase space of TS's.

\begin{acknowledgments}
\paragraph*{Acknowledgments}
We thank Ramon Aguado, Sebastian Bergeret, Francesco Giazotto, Michele Governale, Angela Nigro, and Francesco Romeo for useful discussions.
We acknowledge financial support from the project FIRB-2012-HybridNanoDev (Grant No.~RBFR1236VV).
A.~B. acknowledges funding from the European Union FP7/2007-2013 under REA Grant agreement No.~630925--COHEAT, COST Action MP1209 and STM 2015 of CNR.
\end{acknowledgments}

\appendix
\appendices

\section{Bogololiubov-de~Gennes Hamiltonian}
\label{app:Hamiltonian}

\begin{figure}
\centering
\includegraphics[scale=1,resolution=600]{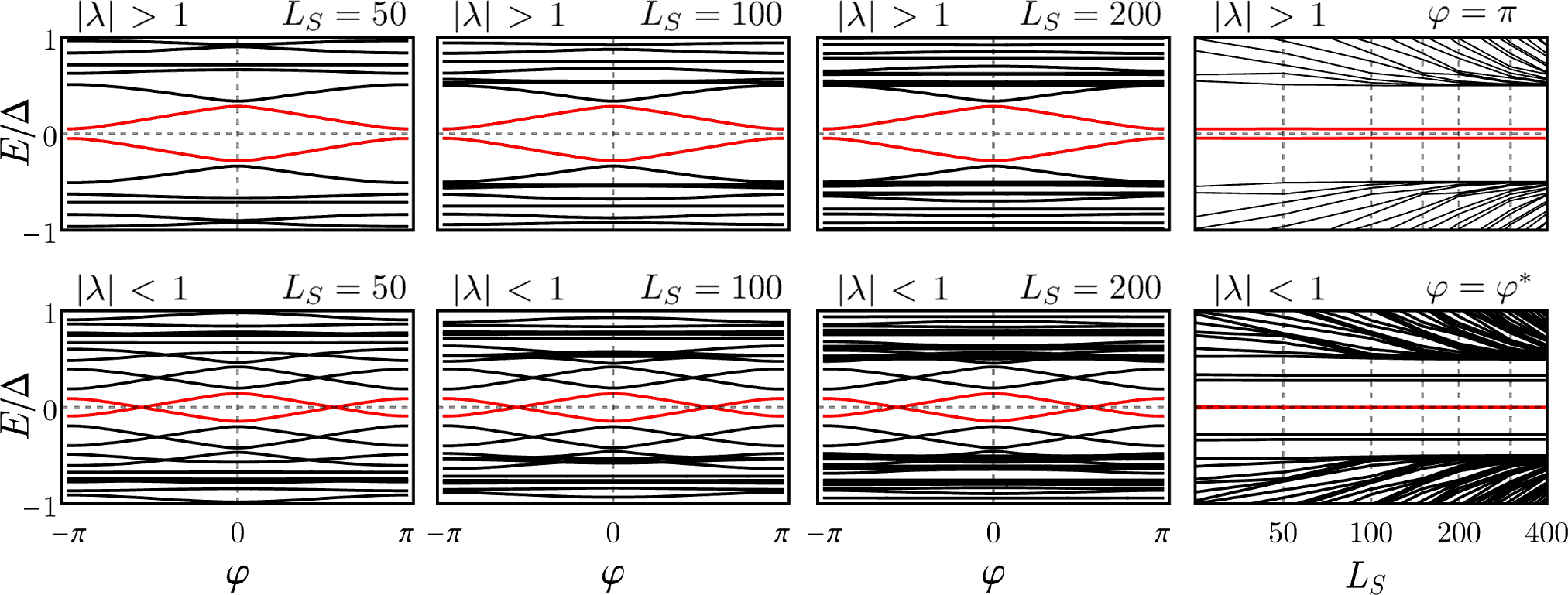}
\caption{
Andreev spectrum of a quantum wire with $L_N=100$, $L_P=50$, $\alpha=0.1t$, $\Delta=0.003t$, $\mu=2t$, for $\mathbf{b}=0$ and $\mathbf{b}/\Delta=(0.3,0.3,0.3)$ 
respectively with $|\lambda|>1$ (above) and $|\lambda|<1$ (below), as a function of the phase $\varphi$ with $L_S=50,100,200$, and as a function of the length $L_S$ of the superconducting leads (rightmost panels) at $\varphi=\pi$ ($|\lambda|>1$, above) and at the gapless point $\varphi^*$ ($|\lambda|<1$ below) respectively.
The LE level (in red) and the gap $\delta_d$ between the LE and the second lowest energy level do not depend on the length $L_S$.
}%
\label{fig:Scaling}
\end{figure}

In general, 
a ballistic Josephson junction\cite{Beenakker2013PRL}
realized by a conventional superconductor in a magnetic field
can be described by a tight-binding BdG Hamiltonian\cite{deGennes1999} with broken 
time-reversal symmetry 
and with an $s$-wave spin-singlet superconducting pairing.
In order to perform actual calculations, we consider specifically the following Hamiltonian
\begin{gather}
\mathcal{H}(\varphi)=
\tfrac12\sum_i
\boldsymbol\Psi_{i}^\dag
\!\cdot\!
\begin{bmatrix}
\mathbf{b}_i\cdot\boldsymbol{\sigma}-\mu\sigma_0&\imath\sigma_y\Delta_i(\varphi)\\
[\imath\sigma_y\Delta_i(\varphi)]^\dagger&-(\mathbf{b}_i\cdot\boldsymbol{\sigma}-\mu\sigma_0)^\intercal
\end{bmatrix}
\!\cdot\!
\boldsymbol\Psi_{i}
\nonumber\\
-
\tfrac12\!\!\!\!\sum_{\langle i,j\rangle\parallel x}\!\!\!\!
\boldsymbol\Psi_{i}^\dag
\!\cdot\!\!
\begin{bmatrix}
t\sigma_0\!+\!\imath\alpha_{ij}\sigma_y&\!0\\
0&\!\!\!\!\!\!-(t\sigma_0\!+\!\imath\alpha_{ij}\sigma_y)
\end{bmatrix}
\!\!\cdot\!
\boldsymbol\Psi_{j}
+\text{h.c.},
\nonumber\\
-
\tfrac12\!\!\!\!\sum_{\langle i,j\rangle\parallel y}\!\!\!\!
\boldsymbol\Psi_{i}^\dag
\!\cdot\!\!
\begin{bmatrix}
t\sigma_0\!+\!\imath\alpha_{ij}\sigma_x&\!0\\
0&\!\!\!\!\!\!-(t\sigma_0\!+\!\imath\alpha_{ij}\sigma_x)
\end{bmatrix}
\!\!\cdot\!
\boldsymbol\Psi_{j}
+\text{h.c.},\!\!\!\!
\label{eq:BdGHamiltonian}
\end{gather}
where $\boldsymbol\Psi_i=(c_{i\up},c_{i\down},c^\dag_{i\up},c^\dag_{i\down})$ is the Nambu spinor, $\sigma_\eta$ the Pauli matrices, $\mu$ the chemical potential, $t$ the hopping parameter, $\mathbf{b}_i=-\mu_B \mathbf{B}_i/\hbar$ the Zeeman magnetic field, $\alpha_{ij}$ the Rashba SOC, and $\Delta_i(\varphi)$ the $s$-wave superconducting order parameter.
We assume uniform magnetic field $\mathbf{b}_i=\mathbf{b}$ and SOC $\alpha_{ij}=\alpha\ge0$ inside the nanostructure (of length $L$ and width $W$ in units of lattice sites); 
$\mathbf{b}_i=\alpha_{ij}=0$ in the superconducting leads; 
$\Delta_i(\varphi)=0$ in the normal region of the nanostructure (of length $L_N$);
$\Delta_i(\varphi)=\Delta$ and $\Delta e^{i\varphi}$ respectively in the two superconducting leads (of length $L_S$) and in the two contiguous proximized regions (of length $L_P$).
The coordinate system is chosen
such that 
the quantization axis is along the $z$ direction, 
the supercurrent flows along the $x$ direction, and the Rashba SOC acts along the $z$ direction, being $\propto(\boldsymbol{\sigma}\times\mathbf{p})_z $.
The BdG Hamiltonian~\ref{eq:BdGHamiltonian} can describe a quantum dot ($L=W=1$ lattice site), a wire ($L>W=1$), or a planar well ($L,W>1$).
Moreover, we assume that, in general, the nanostructure is smaller than the superconducting leads, i.e., $L\ll L_S$.
\Cref{fig:Scaling} 
shows the energy spectrum of a quantum wire ($W=1$) with $L_N=100$, $L_P=50$, $\alpha=0.1t$, $\Delta=0.003t$, $\mu=2t$, and for 
for $\mathbf{b}=0$ and $\mathbf{b}/\Delta=(0.3,0.3,0.3)$ 
as a function of the phase $\varphi$ and of the length of the superconducting leads $L_S$.
The number of energy levels increases as the length of the superconducting leads $L_S$ increases, but the low-energy Andreev spectrum (and in particular the LE level) remains unchanged (see the rightmost panels of 
\cref{fig:Scaling}).
In this case in fact, the degrees of freedom of the superconducting leads may be integrated out, and consequently the low-energy Andreev spectrum can be described by an effective tight-binding Hamiltonian in a finite-dimensional Hilbert space where the superconducting order parameter within the leads is replaced by an effective order parameter at the boundary of the nanostructure\cite{Affleck2000}.
Hence, the low-energy Andreev spectrum does not depend on the ring length.
Therefore in our numerical analysis we consider the limit of infinite ring length.
The analytical results obtained in this work are largely independent on the details of the Hamiltonian~\ref{eq:BdGHamiltonian}, and in particular do not depend on the boundary conditions (open or close) or on the presence of small perturbations (e.g., disorder).

The Andreev spectrum of a confined electron system is a discrete set of energy levels $E_i(\varphi)$ which depends on the gauge-invariant phase difference $\varphi$.
Moreover, the Andreev levels of the Hamiltonian~\ref{eq:BdGHamiltonian} are continuously differentiable functions of the gauge-invariant phase difference $\varphi$.
In fact, the energy levels are the roots of the characteristic polynomial of the effective low-energy Hamiltonian, whose coefficients depend smoothly on the gauge-invariant phase difference $\varphi$.
As shown in Refs.~\onlinecite{Bronshtein1979_Colombini2012}, any polynomial whose roots are real and whose coefficients are smooth functions of a parameter admits roots which are continuously differentiable in that parameter.
Therefore the Andreev levels of the Hamiltonian~\ref{eq:BdGHamiltonian} are continuously differentiable in the phase $\varphi$, for a suitable ordering which removes level-crossing discontinuities.

\Cref{fig:PhaseSpacePfaffianJosephson,fig:JosephsonDerivative} 
have been obtained calculating the Andreev spectrum, the Josephson current, and the Pfaffian of the system described by the Hamiltonian~\ref{eq:BdGHamiltonian} and assuming open boundary conditions.
In particular for the quantum wire 
[\cref{fig:1D,fig:JosephsonDerivative,fig:Majorana}] 
we use
$L_N=100$, $L_P=50$, $W=1$, $\alpha=0.1t$, $\Delta=0.003t$, $\mu=2t$, and $b_y/\Delta=0$, 0.2, 0.3, 0.4, 0.5, and 0.6, which correspond respectively to $\lambda=1.1$, 0.7, 0.2, -0.3, -0.7, and -1.1.
For the planar quantum well 
[\cref{fig:2D}] 
we use instead
and $L_N=20$, $L_P=10$, $W=4$, $\alpha=0.3t$, $\Delta=0.003t$, $\mu=2t$, and $b_y/\Delta=14.0$, 14.4, 14.7, 15.0, 15.2, and 15.6, which correspond respectively to $\lambda=1.2$, 0.4, -0.1, -0.6, -0.9, and -1.5.
Note that the lattice parameters of InAs and InSb nanowires are $\approx 4.5$~\AA, which corresponds to a total length of $\approx 100$~nm for $L=2L_P+L_N=200$ lattice sites.

\section{Phase dependence of the Pfaffian}
\label{app:Pfaffian}

The $\mathbb{Z}_2$ invariant is defined in terms of the Pfaffian\cite{Wimmer2012} of the BdG Hamiltonian in Majorana representation, which is the real number defined as $\mathcal{F}_\varphi=\pf\left[\mathcal{H}(\varphi)\imath\tau_x\right]$.
Since the Hamiltonian is periodic in the order parameter phase $\varphi$, the Pfaffian can be expanded as a Fourier series as
\begin{equation}
\mathcal{F}_\varphi=
A_0+\sum_{n=1}^\infty A_n \cos{\left(n\varphi-\theta_n\right)}=
\sum_{n=-\infty}^\infty C_n e^{\imath n\varphi}
,
\label{eq:PfaffianExpansion0}
\end{equation}
where $A_n$ and $C_n$ are respectively real and complex factors, with $C_0=A_0$ and $C_n=e^{\pm\imath\theta_n}A_n/2$ respectively for $n\lessgtr0$.
The tight-binding BdG Hamiltonian of an $s$-wave TS can be written in Majorana representation as
\begin{align}
\mathcal{H}(\varphi)\imath\tau_x\!&=\frac\imath2\!\!
\begin{bmatrix}
	\imath\sigma_y\Delta_1 \!\!\!\!&&&&&\\[-2mm]
	&\!\!\!\ddots&&&H_0\\[-2mm]
	&&\!\!\!\!\!\imath\sigma_y\Delta_N\!\!\!\\
	&&&\!\!\!\!-\imath\sigma_y\Delta_1^* \!\!\!\!\\[-2mm]
	&\!\!\!-H_0^\intercal&&&\ddots\\[-2mm]
	&&&&&\!\!\!\!\!\!\!\!-\imath\sigma_y\Delta_N^*\\
\end{bmatrix}
\!,
\label{eq:BdGHamiltonianMatrix}
\end{align}
where $H_0$ does not depend on the order parameter, and $N$ is the total number of lattice sites.
Note that $\mathcal{H}(\varphi)\imath\tau_x$ is indeed an antisymmetric matrix, since the terms $\imath\sigma_y\Delta_i$ are antisymmetric.
The Pfaffian of an antisymmetric matrix $A$ of order $2n$ is defined as the polynomial of matrix entries given by
\begin{equation}
\pf(A)=\sum_\pi \sgn(\pi) \, a_{i_1,j_1}a_{i_2,j_2}\ldots a_{i_n,j_n}
,
\label{eq:Pfaffian}
\end{equation}
where the sum is over all permutations in the form
$$
\pi=\begin{pmatrix}1 & 2 & 3&\dots & 2n\\i_1&j_1&i_2&\dots& j_n\end{pmatrix},
$$
and such as $i_k<j_k$ and $i_k<i_{k+1}$ for any index $k$, and where $\sgn(\pi)$ is the sign of the permutation (see, e.g., Ref.~\onlinecite{Wimmer2012}).
The Pfaffian is related to the determinant of antisymmetric matrices by the relation $\pf(A)^2=\det(A)$.
For the properties of permutations, each matrix element of $\mathcal{H}(\varphi)\imath\tau_x$ can appear only once in any of the product in \cref{eq:Pfaffian}.
Moreover, for the additional restriction $i_k<j_k$ on the permutations in \cref{eq:Pfaffian}, one has to consider only products of matrix elements above the diagonal.
As a consequence each term of the Pfaffian can either depend linearly or not depend on the matrix elements $\Delta_i$ and $\Delta_i^*$.
Therefore, one can expand the Pfaffian $\mathcal{F}_\varphi$ as
\begin{equation}
\mathcal{F}_\varphi=
\sum_{p_k,q_k=0}^1
\mathcal{P}^{p_1\cdots p_N}_{q_1\cdots q_N}
\Delta_1^{p_1} 
\ldots \Delta_N^{p_N}
\left(\Delta_1^{q_1}
\ldots \Delta_N^{q_N}\right)^*
,
\label{eq:PfaffianExpansion1}
\end{equation}
where $\mathcal{P}^{p_1\cdots p_N}_{q_1\cdots q_N}$ are complex prefactors which do not depend on the order parameter.
Note that each term $\Delta_1^{p_1}, \ldots, \Delta_N^{p_N}, (\Delta_1^{q_{1}})^*,\ldots, (\Delta_N^{q_N})^*$ of the Hamiltonian appears at most once in \cref{eq:PfaffianExpansion1} according to the properties of the Pfaffian.
Since the Hamiltonian~\ref{eq:BdGHamiltonianMatrix} is invariant up to a global change of the order parameter phase (i.e., $\Delta_i\rightarrow\Delta_i e^{\imath\Omega}$) and we assumed $\Delta_i=\Delta$ and $\Delta_i=\Delta e^{\imath \varphi}$ in the two superconducting leads respectively, the Pfaffian can depend only on the gauge-invariant phase difference $\varphi$ between the leads (or on its multiples $n\varphi$).
Hence, each term of the Pfaffian expansion \cref{eq:PfaffianExpansion1} has to be proportional to $\Delta^{2n}e^{\pm\imath n\varphi}$, with $n$ having its maximum value $n=N$ in the case that $p_k=q_k=1$ for any $k$.
Therefore the Fourier expansion \cref{eq:PfaffianExpansion0} has a finite number of terms, and is given by
\begin{equation}
\mathcal{F}_\varphi=
A_0+\sum_{n=1}^{N} A_n \cos{\left(n\varphi-\theta_n\right)},
\label{eq:PfaffianExpansion2}
\end{equation}
with $A_n\propto\Delta^{2n}$.
If one defines $2C=\max \mathcal{F}_\varphi - \min \mathcal{F}_\varphi$ and $\lambda=(\max \mathcal{F}_\varphi + \min \mathcal{F}_\varphi)/(2C)$, one obtains 
\begin{align}
\mathcal{F}_\varphi&=
C \left( f_\varphi +\lambda \right)
\nonumber\\&=
C\left[a_0+\sum_{n=1}^{N} a_n \cos{\left(n\varphi-\theta_n\right)}+\lambda\right]
,
\label{eq:PfaffianExpansion3}
\end{align}
where $a_0=A_0/C-\lambda$, $a_n=A_n/C$ for $n\ge1$, and $f_\varphi=\mathcal{F}/C-\lambda$.
It follows that $|f_\varphi|\le1$ and thus $\mathcal{F}_\varphi=0$ only if $|\lambda|\le1$.
Since the expansion in \cref{eq:PfaffianExpansion3} has a finite number or nonzero Fourier coefficients, the Pfaffian $\mathcal{F}_\varphi$ is an analytic and periodic function of the phase $\varphi$, and therefore it has a finite number of zeros in the interval $[-\pi,\pi]$.
The same argument can be applied to the derivatives of any order, i.e., the Pfaffian has a finite number of local maxima, minima, and of saddle points (if any), which are indeed isolated points in the interval $[-\pi,\pi]$.

If the superconducting gap $\Delta$ is small with respect to the relevant energy scales of the system, i.e., if $\Delta\ll t$, \cref{eq:PfaffianExpansion3} can be further simplified.
In this case, since $a_n\propto\Delta^{2n}$, higher order Fourier coefficients can be safely neglected, and one obtains $\mathcal{F}_\varphi\approx A_0+A_1 \cos{(\varphi-\theta_1)}$, which leads to 
\cref{eq:PfaffianLimit}, 
i.e., $\mathcal{F}_\varphi\approx C\left[\cos{\left(\varphi-\theta\right)}+\lambda\right]$.
In this approximation, it can be easily shown using trigonometric identities that the coefficients $\lambda$, $\theta$, and $C$ are given by
\begin{align}
&\lambda=
\frac{\mathcal{F}_{\varphi'}+\mathcal{F}_{\varphi'+\pi}}{2C},
\qquad
\tan{\theta}=
\frac{\mathcal{F}_{\nicefrac\pi2}-\mathcal{F}_{-\nicefrac\pi2}}{\mathcal{F}_0-\mathcal{F}_\pi}
,\nonumber\\
&2C=
\sqrt{(\mathcal{F}_0-\mathcal{F}_\pi)^2+(\mathcal{F}_{\nicefrac\pi2}-\mathcal{F}_{-\nicefrac\pi2})^2},
\label{eq:LambdaTheta}
\end{align}
where $\varphi'$ is any angle.
The phase-shift of the Pfaffian $\theta$ in \cref{eq:LambdaTheta} is related to the phase-inversion symmetry in the system.
If the SOC or the magnetic field vanish, the system is invariant under phase-inversion symmetry, i.e., $E_i(\varphi)=E_i(-\varphi)$ and $\mathcal{F}_\varphi=\mathcal{F}_{-\varphi}$, which mandates $\theta=n\pi$ in \cref{eq:LambdaTheta}.
This invariance is related to the magnetic mirror symmetry operator\cite{Lu2015} $M_{xz}\Theta=K$, where $M_{xz}=\imath\sigma_y$ is the spin mirror reflection across the $xz$ plane and $\Theta=-\imath\sigma_y K$ the time-reversal operator.
In general, the only terms in the BdG Hamiltonian~\ref{eq:BdGHamiltonian} which are affected by complex conjugation are the order parameter term, the Rashba SOC $\propto \imath\sigma_x$, and the term in the magnetic field component $b_y$.
However, if $\alpha=0$, the only complex terms of the Hamiltonian~\ref{eq:BdGHamiltonian} are those in the phase $\varphi$ and in the magnetic field component $b_y$.
In this case, since the choice of the quantization axis is arbitrary, one can always choose the $z$ axis in the direction of the applied magnetic field.
On the other hand, if $\mathbf{b}=0$, the only complex terms of the Hamiltonian are those in the phase and the Rashba SOC term $\propto \imath\sigma_x$, which transforms as $(\imath\sigma_x)^*=-\imath\sigma_x$.
In this case, a rotation of the quantization axis around the axis $y$ is sufficient to restore the sign of the SOC term.
Therefore if $\alpha=0$ or $\mathbf{b}=0$ one has $\mathcal{H}(\varphi) =\mathcal{H}(-\varphi)^*$ and, consequently, the energy spectrum and the Pfaffian are symmetric under phase-inversion, i.e., $E_i(\varphi)=E_i(-\varphi)$ and $\mathcal{F}_\varphi=\mathcal{F}_{-\varphi}$.
Therefore if $\alpha=0$ or $\mathbf{b}=0$ one has $\mathcal{F}_\varphi=\mathcal{F}_{-\varphi}$ and \cref{eq:LambdaTheta} further simplifies to
\begin{align}
&\lambda=\frac{\mathcal{F}_{\nicefrac\pi2}}C=\frac{\mathcal{F}_{-\nicefrac\pi2}}C,
\qquad
\theta=
\begin{cases}
0&\mathcal{F}_0\ge\mathcal{F}_\pi\\
\pi&\mathcal{F}_0<\mathcal{F}_\pi
\end{cases}
,
\nonumber\\
&2C=|\mathcal{F}_0-\mathcal{F}_\pi|.
\end{align}
On the other hand, if the magnetic mirror symmetry is broken, i.e., if $\mathcal{H}(\varphi)\neq \mathcal{H}(-\varphi)^*$, the energy spectrum and the Pfaffian may be no longer symmetric under phase-inversion.
In this case, the Fourier coefficients in \cref{eq:PfaffianExpansion0} transforms under complex conjugation as $C_n\rightarrow C_n^*$ which mandates $\theta_n\rightarrow-\theta_n$.
Therefore, the magnetic mirror symmetry $K$ corresponds to the inversion of the phase-shift of the Pfaffian $\theta\rightarrow-\theta$ in 
\cref{eq:PfaffianLimit} 
and, as a consequence, the Pfaffian is no longer invariant under phase-inversion $\varphi\rightarrow-\varphi$ alone, but is still invariant under the more general transformation $\varphi\rightarrow2\theta-\varphi$, i.e., $\mathcal{F}_{\varphi}=\mathcal{F}_{2\theta-\varphi}$.

There is a close relationship between
the topological invariant in 0D (finite-sized system), i.e., the fermion parity of the groundstate $\mathcal{P}_\varphi=\sgn\mathcal{F}_\varphi$, 
and 
the topological invariant in 1D (continuous limit), i.e., the Majorana number\cite{Kitaev2001} $\mathcal{M}$.
In a system with closed boundary conditions, any gapped phase is characterized by the 1D topological invariant 
\begin{align}
\mathcal{M}&=\sgn\{\pf[{\mathcal{H}\imath\tau_x}]\}\nonumber\\
&=\sgn\{\pf{[\widetilde{\mathcal{H}}(0)\imath\tau_x]}\pf{[\widetilde{\mathcal{H}}(\pi)\imath\tau_x]}\}
, 
\label{eq:Pfaffian1D}
\end{align}
where $\widetilde{\mathcal{H}}(k)$ is the Fourier transform of the Hamiltonian $\mathcal{H}$, as shown in Refs.~\onlinecite{Kitaev2001,Zocher2013_Budich2013}.
From \cref{eq:Pfaffian1D} follows immediately that 
the Majorana number $\mathcal{M}$ coincides with the fermion parity in a system with closed boundary conditions.
In a system with open boundary conditions instead, a similar relationship can be obtained by considering the crossover between open and close boundaries via a weak link at the edge of the system.

\section{CPR at zero temperature}
\label{app:Energy}

In systems with particle-hole symmetry, the Andreev spectrum is particle-hole symmetric, and one has
\begin{equation}
\det\left[\mathcal{H}(\varphi)\right]=
\prod_{E_i}\!E_i(\varphi)=
\prod_{E_i\ge0}\!E_i(\varphi)^2.
\label{eq:EnergyExpansionG0}
\end{equation}
We note that the product on the right side of this equation is positive since the total number of negative Andreev levels is even, due to the spin degree of freedom.
Moreover, any zero-energy mode is at least doubly-degenerate, and can be in general expanded as $E_j(\varphi)\propto(\varphi-\varphi^*)^{m_j}$ near the gapless point $\varphi^*$, where $m_j\ge1$ is the multiplicity of the $j$th LE level at zero energy, which gives
\begin{equation}
\det\left[\mathcal{H}(\varphi)\right]\propto
\prod_{j}(\varphi-\varphi^*)^{m_j}
=(\varphi-\varphi^*)^{\sum_j m_j},
\label{eq:EnergyExpansionG1}
\end{equation}
where the index $j$ spans only the modes which have zero-energy at $\varphi=\varphi^*$, i.e., $E_j(\varphi^*)=0$.
Furthermore, for the relationship between the determinant and the Pfaffian one obtains that $\det[\mathcal{H}(\varphi)]=\det[\mathcal{H}(\varphi)\imath\tau_x]=\pf[\mathcal{H}(\varphi)\imath\tau_x]^2$, which yields
\begin{equation}
\mathcal{F}_\varphi\propto
(\varphi-\varphi^*)^{\frac12\sum_j m_j},
\label{eq:EnergyExpansionG}
\end{equation}
where $\mathcal{F}_\varphi=\pf[\mathcal{H}(\varphi)\imath\tau_x]$ is the Pfaffian in Majorana representation.
Therefore, the lowest order $d$ of the non-zero terms in the Pfaffian expansion in the phase near the gapless point is half the total multiplicity, i.e., 
\begin{equation}
2d=\sum_{j=1}^M m_j.
\label{eq:Multiplicity}
\end{equation}
where $M$ is the number of zero-energy modes $E_j(\varphi^*)=0$.
Note that, due to particle-hole symmetry, $M$ is even.
If $\mathcal{F}_{\varphi^*}^{\prime}\neq0$, the first derivative of the Pfaffian is non-zero ($d=1$) and, as a consequence, the fermion parity 
$\mathcal{P}_\varphi=\sgn\mathcal{F}_\varphi$ 
changes at the gapless point.
In this case \cref{eq:Multiplicity} and particle-hole symmetry mandates that there can be only two doubly-degenerate LE levels $E_\pm(\varphi^*)=0$ with multiplicities $m_\pm=1$, i.e., which have a linear phase-dispersion $E_\pm(\varphi)\propto(\varphi-\varphi^*)$.
This mandates, as we will show, a discontinuous drop in the Josephson CPR\@.
A necessary condition for this case ($\mathcal{F}_{\varphi^*}^{\prime}\neq0$ and $d=1$) is the lifting of the spin degeneracy, due to the broken 
time-reversal symmetry.
If $\mathcal{F}_{\varphi^*}^{\prime}=0$ ($d>1$) instead, 
there is no constraint in general on the phase-dispersion of zero-energy modes, which may be linear ($m_j=1$) or non-linear ($m_j>1$).
For example, if $d=2$, one can satisfy \cref{eq:Multiplicity} 
with $M=4$ particle-hole and spin degenerate modes with linear dispersion $m_1=m_2=m_3=m_4=1$, 
or
with $M=2$ particle-hole symmetric modes with parabolic dispersion $m_1=m_2=2$ if 
time-reversal symmetry 
is broken.
Therefore in the limit points where $\mathcal{F}_{\varphi^*}=\mathcal{F}_{\varphi^*}^{\prime}=0$ the CPR may or may not be discontinuous, respectively if the zero-energy modes have a linear ($m_j=1$) or non-linear ($m_j>1$) phase-dispersion.

\subsubsection{Broken time-reversal symmetry}

In the case of broken 
time-reversal symmetry 
and if $\mathcal{F}_{\varphi^*}^{\prime}\neq0$, one has that $d=1$ in \cref{eq:Multiplicity} and, as a consequence of particle-hole degeneracy, there can be only two degenerate zero-energy modes $E_\pm(\varphi^*)=0$ with multiplicity $m_\pm=1$.
Therefore one can factorize out the doubly-degenerate LE Andreev levels $E_\pm(\varphi)$ in \cref{eq:EnergyExpansionG0} as
\begin{equation}
\det\left[\mathcal{H}(\varphi)\right]=|E_\pm(\varphi)|^2
\prod_{E_i>E_+}\!E_i(\varphi)^2,
\label{eq:EnergyExpansion0}
\end{equation}
where $E_i(\varphi)$ are the Andreev levels with higher energy.
Again for the properties of the determinant and of the Pfaffian one obtains
\begin{equation}
E_\pm(\varphi)=\pm
\frac{|\mathcal{F}_\varphi|}{\chi_\varphi}
,
\label{eq:EnergyExpansion1}
\end{equation}
where $\chi_\varphi=\prod_{E_i>E_+}E_i(\varphi)>0$ is the product of positive Andreev levels with energy greater than the LE level.
Since $\partial_\varphi|\mathcal{F}_\varphi|=(\sgn{\mathcal{F}_\varphi}) \mathcal{F}'_\varphi$ and 
$\sgn{\mathcal{F}_\varphi}=\mathcal{P}_\varphi$ 
is the fermion parity, it follows that
\begin{equation}
\partial_\varphi E_\pm(\varphi)=\pm
\mathcal{P}_\varphi
\left(
\frac{\mathcal{F}_\varphi^{\prime}}
{\chi_\varphi}
-
\frac{\chi'_\varphi\mathcal{F}_\varphi}{\chi_\varphi^2}
\right)
,
\label{eq:DispersionExpansion1}
\end{equation}
where ${\chi'_\varphi}=\partial_\varphi{\chi_\varphi}$ and $\mathcal{F}_\varphi^{\prime}=\partial_\varphi \mathcal{F}_\varphi$.
The contribution to the current of the LE level can be obtained directly from \cref{eq:DispersionExpansion1}, since $I_{le}(\varphi)=(e/\hbar)\partial_\varphi E_-(\varphi)$.
This implies that a change of the fermion parity at the gapless point mandates a sign-change of the LE level contribution to the current.
Moreover, at the gapless point one has $E_\pm(\varphi^*)=\mathcal{F}_{\varphi^*}=0$, the second term in \cref{eq:DispersionExpansion1} vanishes and one obtains
\begin{equation}
|\partial_\varphi E_\pm(\varphi^*)|
=
\frac{|\mathcal{F}_{\varphi^*}^{\prime}|}
{\chi_{\varphi^*}}
,
\label{eq:DispersionExpansion}
\end{equation}
where $\chi_{\varphi^*}=\prod_{E_i>0} E_i(\varphi^*)>0$
is the product of positive energy levels at $\varphi^*$.
The quantity $\chi_{\varphi^*}$ can be related to the pseudo-determinant, i.e., the product of non-zero eigenvalues of the Hamiltonian $\pdet{[\mathcal{H}(\varphi^*)]}=\prod_{E_i\neq0} E_i(\varphi^*)$.
In the case considered ($d=1$) one has $\prod_{E_i\neq0} E_i(\varphi^*)=-\prod_{E_i>0} E_i(\varphi^*)^2$, where the minus sign comes from the fact that there are $2N-1$ negative energy levels $E_i(\varphi^*)<0$.
Therefore one obtains $\chi_{\varphi^*}=\sqrt{|\pdet{[\mathcal{H}(\varphi^*)]}|}$.
In this case ($d=1$, only two zero-energy levels at $\varphi^*$) the pseudo-determinant can be calculated using the identity $\pdet{[\mathcal{H}(\varphi^*)]}=\lim_{\varepsilon\rightarrow0}\varepsilon^{-2}\det{[\mathcal{H}(\varphi^*)+\varepsilon\mathbbm{1}]}$.
\Cref{eq:DispersionExpansion} leads to 
\cref{eq:Discontinuity} 
by noting that 
the level crossing mandates a
discontinuous drop of the Josephson current given by $\Delta I(\varphi^*)=-2(e/\hbar)|\partial_\varphi E_\pm(\varphi^*)|$.
Moreover, in this case the LE Andreev level in the neighborhood of the gapless point $\varphi^*$ is linear in the phase $\varphi$ for $\mathcal{F}_{\varphi^*}^{\prime}\neq0$, and can be expanded at the first order as
\begin{equation}
E_\pm(\varphi)\approx\pm
\mathcal{P}_\varphi
\frac{|\mathcal{F}_{\varphi^*}^{\prime}|}{\chi_{\varphi^*}}
(\varphi-\varphi^*),
\label{eq:LowestAndreevLevel1stOrder}
\end{equation}
where the fermion parity 
$\mathcal{P}_\varphi$
changes its sign at the gapless point $\varphi^*$.
Therefore, discontinuities in the CPR correspond to fermion parity transitions in systems with broken time-reversal symmetry.
For the same reasons, 
discontinuities of the current with respect to any parameter $\nu$ of the Hamiltonian, e.g., magnetic field, chemical potential, or SOC\@, correspond to fermion parity transitions in these systems.

\subsubsection{Unbroken time-reversal symmetry}

In the case that 
time-reversal symmetry 
is unbroken, any zero-energy mode is at least four-fold degenerate due to particle-hole and spin degeneracy.
Therefore, if $\mathcal{F}_{\varphi^*}=0$, there exist at least four energy levels $E_j(\varphi^*)=0$ with multiplicity $m_j\ge1$ and, as a consequence of \cref{eq:Multiplicity}, the total multiplicity is $2d=\sum_j m_j =4n\ge4$.
Hence, the lowest order $d$ of the Pfaffian expansion in the phase near the gapless point is even and therefore $\mathcal{F}_{\varphi^*}^{\prime}=0$.
In this case, the CPR may still be discontinuous, but these discontinuities do not correspond to any topological phase transition, since the Pfaffian $\mathcal{F}_\varphi\propto(\varphi-\varphi^*)^d$ do not change its sign at the gapless point (since $d$ is even).
This is in agreement with the fact that an $s$-wave superconductor is topologically trivial if the 
time-reversal symmetry 
is unbroken.

In the simplest case where no additional degeneracy is present and the multiplicity of each of the four-fold degenerate levels is $m_j=1$, one has $2d=4$ and consequently \cref{eq:EnergyExpansion0} becomes
\begin{equation}
\det\left[\mathcal{H}(\varphi)\right]=|E_\pm(\varphi)|^4
\prod_{E_i>E_+}\!\!E_i(\varphi)^2.
\end{equation}
Moreover, using again the fact that $\det\left[\mathcal{H}(\varphi)\right]=\mathcal{F}_\varphi^{\,2}$ one obtains
\begin{equation}
E_\pm(\varphi)=\pm
\sqrt{\frac{|\mathcal{F}_\varphi|}{\chi_\varphi}}
,
\end{equation}
while the phase-derivative of the LE level is given by
\begin{equation}
\partial_\varphi E_\pm(\varphi)=
\pm
\frac{1}{2\sqrt{\chi_\varphi|\mathcal{F}_\varphi|}}
\left[\!
\mathcal{P}_\varphi
\mathcal{F}^{\prime}_\varphi
-
|\mathcal{F}_\varphi|\frac{\chi'_\varphi}
{\chi_\varphi}
\!\right]\!
,
\label{eq:DispersionExpansionTrivial1}
\end{equation}
where $\mathcal{F}_{\varphi^*}=\mathcal{F}^{\prime}_{\varphi^*}=0$ at the gapless point ($d=2$).
Since the Josephson current cannot diverge, the derivative of the LE energy in \cref{eq:DispersionExpansionTrivial1} is finite at $\varphi^*$.
In the limit of small gap $\Delta\ll t$, one has $\mathcal{F}_\varphi=C\left[\cos{\left(\varphi-\theta\right)}+\lambda\right]$ 
[cf.~\cref{eq:PfaffianLimit}], 
and therefore since $\mathcal{F}_{\varphi^*}=\mathcal{F}^{\prime}_{\varphi^*}=0$ at the gapless point, one must have $\lambda=\pm1$.
Moreover for a vanishing magnetic field (unbroken 
time-reversal symmetry) 
one has that $\theta=n\pi$, and therefore in this case the gap closes at $\varphi^*=n\pi$ for $\lambda=\pm1$.
Finally, in the limit $\varphi\rightarrow\varphi^*$, \cref{eq:DispersionExpansionTrivial1} yields
\begin{equation}
\partial_\varphi E_\pm(\varphi^*)\approx\pm\sqrt{\frac{C}{2\,\chi_{\varphi^*}}}.
\label{eq:DispersionExpansionTrivial}
\end{equation}
Hence, in this case the CPR may still exhibit a single discontinuity at the gapless point $\varphi^*=n\pi$ due to the presence of the four-fold degenerate zero-energy mode, which is given by $\Delta I(\varphi^*)=-{4e}/{\hbar}|\partial_\varphi E_\pm(\varphi^*)|$, and which do not correspond to any topological phase transition.

\section{CPR at finite temperature}
\label{app:Temperature}

At finite temperature, the Josephson current in 
\cref{eq:Josephson} 
can be rewritten using the particle-hole symmetry and the identity $f(x)-f(-x)=-\tanh(x/2)$, which yield
\begin{equation}
I(\varphi)=
-\frac{e}{\hbar}\sum_{E_i\ge0} 
\tanh\left[\frac{E_i(\varphi)}{2k_B T}\right]
\partial_\varphi E_i(\varphi).
\label{eq:JosephsonTanh}
\end{equation}
In the case of broken 
time-reversal symmetry, 
any discontinuity in the CPR which corresponds to a topological phase transition is given by the contribution of a doubly-degenerate ($d=1$) Andreev level 
(cf.~\cref{app:Energy}).
The contribution of the LE level $E_\pm(\varphi)$ can be considered separately, and one can evaluate the CPR as the sum of the current contributions $I_{le}(\varphi)$ and $I_{he}(\varphi)$, respectively of the LE level $E_+(\varphi)\ge0$ and of higher energy levels $E_i(\varphi)>\delta_d$, where $\delta_d$ is the energy gap between the lowest and the higher Andreev levels.
At low temperatures $T\ll T_d=\delta_d/k_B$, for any level $E_i(\varphi)>\delta_d$ one has $\tanh[E_i(\varphi)/(2k_B T)]\approx1$ in \cref{eq:JosephsonTanh}, and therefore the contribution to the Josephson current of higher energy levels does not depend on temperature, and can be approximated as a linear function of the phase $\varphi$, i.e., $I_{he}(\varphi)\approx I_{he}(\varphi^*)+I_{he}^\prime(\varphi^*)(\varphi-\varphi^*)$ near the closing gap $\varphi^*$.
On the other hand, the contribution of the LE Andreev level can be obtained from \cref{eq:LowestAndreevLevel1stOrder} which gives $E_+(\varphi)\approx-{\hbar}/{(2e)}\Delta I(\varphi^*)(\varphi-\varphi^*)$, and therefore
\begin{equation}
I_{le}(\varphi)\approx
\frac{\Delta I(\varphi^*)}{2}
\tanh\left[-\frac{\hbar}{e}
\frac{\Delta I(\varphi^*)(\varphi-\varphi^*)}{4k_B T}\right],
\label{eq:JosephsonLE}
\end{equation}
which leads to 
\cref{eq:JosephsonSmooth}.
Furthermore, the first order phase-derivative of the current near the closing of the particle-hole gap yields
\begin{equation}
\partial_\varphi I(\varphi)\!\approx\!
-\frac{\hbar}{e}\frac{\Delta I(\varphi^*)^2}{8k_B T}
\sech^2\!\!\left[-\frac{\hbar}{e}
\frac{\Delta I(\varphi^*)(\varphi-\varphi^*)}{4k_B T}\!\right]\!\!,
\!
\label{eq:JosephsonSech}
\end{equation}
where we neglect the contribution of higher energy Andreev levels.
It follows that the phase-derivative of the Josephson current at low temperatures is $\partial_\varphi I(\varphi^*)\approx-({\hbar}/{e})\Delta I(\varphi^*)^2/(8k_B T)$ at the gapless point, and diverges for $T\rightarrow 0$.
For this reason the scaling of the phase-derivative in the limit $T\rightarrow0$ can be used to directly measure the discontinuous drop $\Delta I(\varphi)$.

We notice that the fundamental premise for the observability of the effects discussed in this work is the detectability of the discrete nature of the subgap Andreev levels of the Josephson junction.
This means that, in the ballistic limit, the average energy spacing between Andreev levels in the low energy (subgap) spectrum should be larger than the other relevant energy scales of the system.
In particular, the temperature should be sufficiently small, i.e., $T\ll T_d=\delta_d/k_B$,
and disorder sufficiently weak, i.e., at least $\hbar/\tau\ll\delta_d$ with $\tau$ the scattering time in the nanostructure.
Numerical calculations have confirmed that the presence of weak disorder
does not affect the relation between the topological phase transitions and the discontinuities in the CPR, as long as the time-reversal symmetry is broken.

\section{Majorana polarization}
\label{app:Majorana}

The total Majorana polarization\cite{Sticlet2012} at any lattice site $r_i$ and at energy $E$ is defined as the complex number
\begin{equation}
p(r_i,E)=\sum_{j}2\left(u_{i\down} v_{i\down}^{*}+u_{i\up} v_{i\up}^{*}\right)_j\delta(E-E_j),
\label{eq:MajoranaPolarization}
\end{equation}
where the sum is over all eigenstates $\psi_j$ with energy $E_j$, $\delta(E)$ the Dirac delta function, and where $u_{is}$ and $v_{is}$ are the particle-hole representation of the eigenstate $\psi_j=\sum_i u_{i\up} c_{i\up}+ u_{i\down} c_{i\down}+ v_{i\up} c^\dag_{i\up}+ v_\down c^\dag_{i\down}$.
Note that the definition in \cref{eq:MajoranaPolarization} is consistent with our choice of the Nambu basis $\boldsymbol\Psi_i=(c_{i\up},c_{i\down},c^\dag_{i\up},c^\dag_{i\down})$.
For a pure particle state ($v_{i s}=0$) or pure hole ($u_{i s}=0$) one has $2(u_{i\down} v_{i\down}^{*}+u_{i\up} v_{i\up}^{*})=0$ while for Majorana edge states localized at the lattice site $r_i$ in the form
\begin{align}
\gamma_{is}^+=\frac1{\sqrt{2}} \left(e^{\imath\varphi/2} c^\dag_{is}+ e^{-\imath\varphi/2} c_{is}\right),
\nonumber\\
\gamma_{is}^-=\frac\imath {\sqrt{2}} \left(e^{\imath\varphi/2} c^\dag_{is}-e^{-\imath\varphi/2} c_{is}\right),
\end{align}
one has $2(u_{i\down} v_{i\down}^{*}+u_{i\up} v_{i\up}^{*})=\pm e^{\imath\varphi}$, where the sign corresponds to the two different flavors of Majorana states.
Any linear superposition of particle and hole states with $u_{i s}\neq0$ and $v_{i s}\neq0$ can be written as a superposition of two orthogonal Majorana states $a_+ \gamma_{is}^+ + a_-\gamma_{is}^-$.
Such particle-hole modes have in general a finite Majorana polarization $2(u_{i\down} v_{i\down}^{*}+u_{i\up} v_{i\up}^{*})\neq0$.

\begin{figure}
\centering
\includegraphics[scale=1,resolution=600]{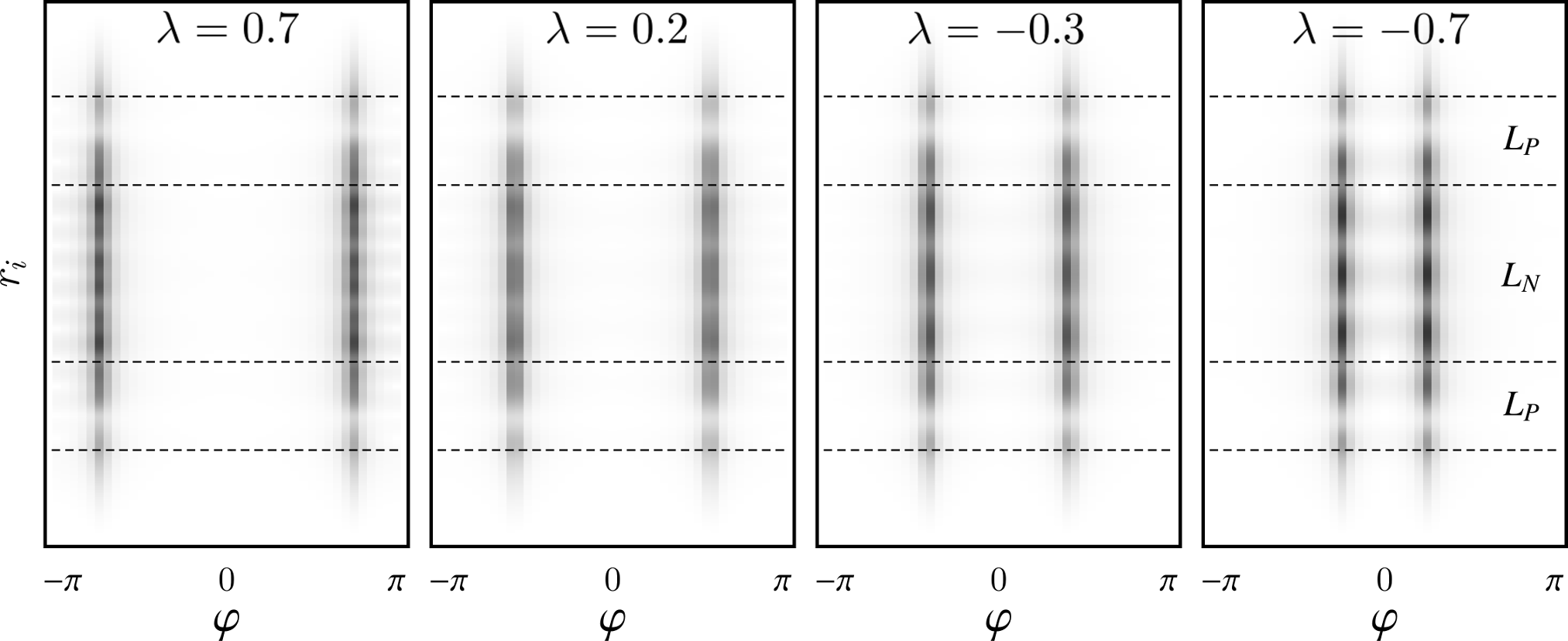}
\caption{
Majorana polarization $|p(r_i,0)|^2$ at zero energy for $|\lambda|<1$ in a wire as a function of the gauge-invariant phase $\varphi$ and lattice site $r_i$.
}%
\label{fig:Majorana}
\end{figure}

At the gapless points, where the fermion parity changes its sign, the presence of zero-energy modes can be revealed by the Majorana polarization at zero energy 
shown in \cref{fig:Majorana} 
as a function of the gauge-invariant phase $\varphi$ and lattice site $r_i$.
The Majorana polarization is finite at gapless points $\varphi^*_\pm$ inside the nanostructure.
This entails that such zero-energy modes are not purely hole nor particle-like, but a superposition of orthogonal Majorana states.
Note that these zero-energy modes are not localized at the edges, but are delocalized along the whole nanostructure.

\end{document}